\documentclass[preprint,showpacs,preprintnumbers]{revtex4}
\usepackage{graphicx}
\usepackage{dcolumn}
\usepackage{bm}

\input{tcilatex}

\begin{document}

\preprint{}
\title{Andreev reflection and enhanced subgap conductance in
NbN/Au/InGaAs-InP junctions}
\author{I. E. Batov$^{1,2}$}
\email{i.batov@fz-juelich.de}
\author{Th. Sch\"apers$^2$}
\author{A. A. Golubov$^3$}
\author{A. V. Ustinov$^1$}

\affiliation{$^1$Physikalisches Institut III, Universit\"at
Erlangen-N\"urnberg, Erwin-Rommel-Str. 1, 91058 Erlangen, Germany\\
$^2$Institut f\"ur Schichten und Grenzfl\"achen and\\
cni~-~Center of Nanoelectronic Systems for Information Technology,
Forschungszentrum J\"ulich, 52425 J\"ulich,Germany\\
$^3$Faculty Faculty of Applied Physics, University of Twente, P.
O. Box 217, 7500 AE Enschede, The Netherlands}
\date{\today }

\begin{abstract}
We report on the fabrication of highly transparent
superconductor/normal metal/two-dimensional electron gas junctions
formed by a superconducting NbN electrode, a thin (10nm) Au
interlayer, and a two-dimensional electron gas in a InGaAs/InP
heterostructure. High junction transparency has been achieved by
exploiting of a newly developed process of Au/NbN evaporation and
rapid annealing at 400$^\circ$C. This allowed us to observe for
the first time a decrease in the differential resistance with
pronounced double-dip structure within the superconducting energy
gap in superconductor-2DEG proximity systems. The effect of a
magnetic field perpendicular to the plane of the 2DEG on the
differential resistance of the interface was studied. It has been
found that the reduced subgap resistance remains in high magnetic
fields. Zero-field data are analyzed within the previously
established quasiclassical model for the proximity effect.

\end{abstract}

\maketitle



\section{Introduction}

The study of hybrid systems consisting of superconductors (S) in contact
with a two-dimensional electron gas (2DEG) has attracted considerable
interest in recent years. Various new effects arising due to the Andreev
reflection at the S/2DEG interface were studied in such systems, both
theoretically and experimentally \cite%
{Neurohr96,Takagaki98,Asano00,Asano00a,Chtchelkatchev01,
Hoppe00,Zuelicke01,Tkachov03,Zhou98,Moore99,Uhlisch00,Takayanagi02}. By
taking phase-coherent Andreev reflection into account, the oscillations in
the conductance of a S/2DEG junction in magnetic field have been predicted
which are based on an Aharonov-Bohm-type interference effect \cite%
{Takagaki98,Asano00,Asano00a,Chtchelkatchev01}. The current transport across
the S/2DEG interface in the quantum Hall effect regime was investigated by
Hoppe, Z\"ulicke and Sch\"on \cite{Hoppe00,Zuelicke01}. They could show that
bound states are formed at the S/2DEG interface which are the coherent
superposition of electron and hole edge excitations \cite{Hoppe00}. Very
recently, the magneto-tunneling spectrum and the thermal conductance have
been studied in a superconductor/quasi-one-dimensional semiconductor
structure \cite{Tkachov03}. The transport across a
superconductor/semiconductor interface in the Hall geometry was discussed in
\cite{Takagaki98,Zhou98}. In \cite{Zhou98}, it is predicted that the Hall
voltage is significantly suppressed near the interface compared to its
normal metal value.

Experimental observation of these new phenomena is a challenging
problem because in most cases the high probability of Andreev
reflection at the interface is essential. Since Andreev reflection
is a two-particle process \cite{Andreev64}, it is strongly
affected by the transmissivity at the S/2DEG interface. Therefore,
a serious effort has to be made to achieve a high junction
transparency in order to access the novel transport regimes of
interest. Recently, Takayanagi et al. \cite{Takayanagi02} reported
the succesful preparation of good ohmic contacts between
NbN/AuGeNi electrodes and a 2DEG in an AlGaAs/GaAs
heterostructure.

In this work, we report on the fabrication of highly transparent
superconductor/2DEG junctions formed by superconducting NbN
electrodes and a 2DEG in a InGaAs/InP heterostructure. High
junction transparency has been achieved by the use of a thin
(10~nm) Au interlayer between NbN and the heterostructure. This
achievement allowed us to observe for the first time a decrease in
the differential resistance with pronounced double-dip structure
within the superconducting energy gap and to measure the
dependence of this structure on the magnetic field. Previously,
experiments on transparent S/2DEG junctions (in diffusive limit)
revealed a peculiar non-monotonic voltage dependence of the
differential resistance with a maximum at zero bias, referred to
as  "reentrant" resistance
\cite{Charlat96,VolkovAllsopp96,GolubovWilhelm97,Nazarov96,Belzig99}.
In diffusive S/2DEG contacts with Schottky barriers, a large and
narrow peak in the differential conductance of the junctions
around zero bias voltage was observed
\cite{Kastalsky91,Magnee94,Quirion02}. The latter effect, known as
"reflectionless tunneling", has been studied theoretically in
Refs.
\cite{Zaitsev90,Lambert91,Wees92,Takane92,Beenakker92,VolkovKlapwijk92,VolkovZaitsev93,
Volkov93,Hekking93,Beenakker94,MelsenBeenakker96,Lesovik97,VolkovTakayanagi97,Bezuglyi00}.
Below we will argue that the new double-dip structure in the
differential resistance detected in our measurements is related to
the transport in SN-2DEG contacts in a ballistic regime
\cite{Neurohr96}. Experimental data for zero magnetic field will
be analyzed in our work within the quasiclassical model for the
proximity effect developed in \cite{Neurohr96,Golubov96}.

\section{Experimental}

Our experiments have been performed with strained InGaAs/InP
heterostructures fabricated by metal organic vapor phase epitaxy (MOVPE) on
a semi-insulating InP substrate. The InGaAs/InP heterostructures consist of
a 400~nm thick InP buffer, a 10~nm thick (Si)-InP dopant layer ($N_d=4.2
\times 10^{17} \mathrm{cm}^{-3}$), a 20~nm thick InP spacer, a 10~nm thick In%
$_{0.77}$Ga$_{0.23}$As layer containing the 2DEG, and a 150~nm thick In$%
_{0.53}$Ga$_{0.47}$As cap. The use of InGaAs/InP heterostructures
with a strained In$_{0.77}$Ga$_{0.23}$As layer allows one to
achieve a high mobility in the 2DEG owing to a low effective
electron mass of $m^* = 0.036~m_e$ and a reduced contribution of
alloy scattering. Shubnikov-de Haas and Hall effect measurements
of our 2DEG structures revealed a carrier concentration of $7
\times 10^{11}~\mathrm{cm}^{-2}$ and a mobility of about 250000
cm$^2$/Vs at 0.3~K. From these values a Fermi energy $E_f$ of
37~meV and a transport mean free path of 3.6 $\mu$m were
estimated.

A schematic of the sample layout is shown in Fig.~1. The semiconductor mesa
was defined by electron beam lithography and reactive ion etching (RIE)
using CH$_4$/H$_2$ gas mixture. We prepared two different types of
electrodes to the two-dimensional electron gas in a InGaAs/InP
heterostructure. The electrodes of the first type were 100~nm thick NbN
layers deposited by dc magnetron sputtering. The NbN electrodes are
contacted at the mesa sidewalls to the 2DEG. Directly prior to the
deposition of NbN, the semiconductor surface was cleaned by an Ar plasma, in
order to remove residual atoms at the surface. At the final processing step,
the geometry of the NbN electrodes were defined using a lift-off technique.
The electrodes of the second type were prepared by a modified two-step
process. First, after the cleaning of the surface of the etched InGaAs/InP
heterostructure by Ar-plasma, a thin (10~nm) Au interlayer was deposited in
situ by dc magnetron sputtering followed by the deposition of the NbN
electrode. Subsequently, we employed a rapid ($\approx$ 10~sec) annealing
step. The temperature of annealing was 400$^\circ$C.

For the electrical characterization a three-terminal measurement scheme has
been employed where a small ac current is superimposed to a dc bias current.
The differential resistance is obtained by detecting the ac voltage by a
lock-in amplifier. The ac excitation current was 10~nA. The measurements
were performed at temperatures down to 0.3~K in a He-3 cryostat equipped
with a superconducting solenoid with a magnetic field up to 10~T.

\section{Experimental Results}

In Fig.~\ref{Fig2}, the differential resistance $dV/dI$ as a function of the
bias voltage measured at different temperatures is shown for a 10~$\mu$m
wide NbN/2DEG interface. It is seen that at low temperatures, a pronounced
resistance peak occurs at zero bias voltage, indicating a strong barrier at
the NbN/2DEG interface. Measurements show that the height of the resistance
peak does not depend on temperature in the low temperature range $T<1.5$~K.
If the DC bias voltage approaches a value of 1.5~mV, a minimum in the
differential resistance is observed. The position of this resistance minimum
coincides with the superconducting energy gap determined from the critical
temperature of superconductor $T_c$. At voltages $V_{dc} > 4$~meV a constant
differential resistance is found. With increasing the temperature the
resistance peak at zero bias as well as the minima at about $\pm1.5$~mV
become weaker and disappear completely at temperatures above $T_c \approx 10$%
~K. The interface resistances of the NbN/InGaAs-InP interfaces are
relatively high due to a high barrier at the interface. In order to improve
the contact characteristics, we modified the method of preparation of the
S/2DEG interfaces of the structures as was discussed in the previous
section. By using a thin gold interlayer, we succeeded to obtain a
considerably smaller interface resistance.

Figure~\ref{Fig3} shows the differential resistance of a $5~\mu$m wide
NbN/Au/2DEG interface as a function of the voltage drop at different
temperatures. It is seen that a decrease in the differential resistance is
observed within the range of voltages $\pm 1$~mV at temperatures below 6~K.
This result is a characteristic of the transport across a junction with high
probability of Andreev reflection, resulting in excess current at low bias
\cite{Neurohr96,Golubov96}. At zero dc bias voltage, the differential
resistance exhibits a minimum. At 0.5~K, the zero bias resistance dip
reaches a value of about 15 \% of the normal state resistance $R_N$. At low
temperatures ($<2$~K), in the voltage dependence of the differential
resistance, we observed two shoulders, both symmetric in voltage. As the
temperature is increased, the double dip structure in the differential
resistance progressively disappears. At temperatures higher than 2~K, the
feature at small voltages is suppressed and only a broad resistance dip
within the range of 1~mV remains. Finally, at temperatures about 6~K almost
constant differential resistance is measured. At this temperature the sample
is effectively normal conductive, since the temperature is close to $T_c$.

Fig.~\ref{Fig4} demonstrates the effect of a magnetic field perpendicular to
the plane of the 2DEG on the differential resistance of the NbN/Au/2DEG
interface. The differential resistance is plotted versus dc bias current $%
I_{dc}$ and is normalized to the resistance value $R_i$ measured at $1.5~\mu$%
A. The measurements were performed at 0.5~K. It is seen that with
increasing magnetic field $B$ the reduction of the resistance at
zero bias as related to its value at $I_{dc}=1.5~\mu$A becomes
smaller. The double-dip structure
occurs on the scale of about several 100~mT and is still clearly seen at $%
B=0.9$~T. At high fields, the features are no longer well
resolved, however a reduced zero-bias resistance remains even at
magnetic fields of about 3~T. Note that in our samples, the onset
of the Shubnikov-de Haas oscillations in the 2DEG is observed at a
magnetic field of about 0.25~T. Therefore, our experiments
indicate that the subgap conductance enhancement due to Andreev
reflection is preserved in the regime of high quantizing magnetic
fields.

\section{Discussion and Comparison with Theory}

Detailed interpretation of the I-V curves in magnetic field
requires separate study and will be reported elsewhere. Here we
shall discuss the experimental data in zero field, in order to
demonstrate that the regime of highly-transparent S/2DEG
interfaces was achieved in our junctions.

First we analyze the experimental data using the well-established Blonder,
Tinkham, and Klapwijk (BTK) model for the current transport \cite{Blonder82}%
. Although the BTK model does not include effects of disorder and an
intermediate non-superconducting layer at the interface \cite{Neurohr96}, it
is widely applied to both ballistic and diffusive systems, in order to
obtain an estimate of the junction transmissivity. In the BTK model, a
potential barrier at the interface is approximated by a $\delta$-function
potential. A transmission coefficient $T_N$ in the normal state is given by $%
T_N=1/(1+Z^2)$ where $Z$ is the dimensionless parameter characterizing the
potential barrier strength \cite{Blonder82}. Following \cite{Gao93}, from
the drop of the differential resistance at zero bias compared to the large
bias case (experimental curve at $T= 0.5$~K, Fig. 3), a barrier strength $%
Z=0.5$ for the NbN/Au/InGaAs-InP junction is obtained. The normal-state
transmission coefficient $T_N$ calculated from $Z$-value is 0.8, which is
considerably higher than the value of $T_N=0.2$ obtained for NbN/InGaAs-InP
junctions.

Golubov {\it et al.} \cite{Neurohr96,Golubov96} extended the BTK
model by taking into account the presence of a non-superconducting
N layer at the interface. The structure therefore may be
represented as an SN-2DEG junction. The S and N layers are assumed
to be in the dirty limit, while the 2DEG channel is in the clean
limit \cite{Neurohr96}. The N-2DEG interface is simulated in the
model by the BTK $Z$-factor.

The equilibrium state of the S and N layers is described by the
angle-averaged Green's functions $G$ and $F$, which are obtained from the
Usadel equations \cite{Belzig99}. By formally introducing a complex angle $%
\theta(\epsilon,x)$ with $G(\epsilon,x)=\cos \theta(\epsilon,x)$ and $F=
\sin \theta(\epsilon,x)$, the Usadel equations are written as \cite%
{Neurohr96}:
\begin{equation}
\xi_{S,N}^2 \theta_{S,N}^{\prime \prime} (x)+ i \epsilon \sin \theta_{S,N}
(x)+ \widetilde{\Delta}_{S,N} \cos \theta_{S,N}(x)=0 \   \label{eq-1}
\end{equation}
where $\epsilon =E/\pi k_B T_c$ and $\widetilde{\Delta}_{S,N}=\Delta_{S,N}/%
\pi k_B T_c$ are the normalized energy and pair potential, and $T_c$ is the
critical temperature of the superconductor. The coherence lengths $\xi_{S,N}$
are given by $\xi_{S,N}=(\hbar D_{S,N}/2\pi k_B T_c)^{1/2}$ where $D_S$ and $%
D_N$ are the diffusion coefficients in the superconductor and normal layers,
respectively.

The pair potential $\widetilde{\Delta}_{S}$ in the superconductor is
determined by the self-consistency equation
\begin{equation}
\widetilde{\Delta}_S (x)\ln \frac{T}{T_c} + 2 \frac{T}{T_c} \sum_{\omega_n
>0} \left[\frac{\widetilde{\Delta}_S(x)}{\omega_n}- \sin \theta_S \left( i
\omega_n,x\right)\right]=0  \label{eq-2}
\end{equation}
where $\omega_n= (2n+1)T/T_c$ are the normalized Matsubara frequencies. In
the normal conductor, the pair potential $\Delta_N$ is assumed to be zero.
The quasiparticle density of states (DOS) in terms of the proximity angle $%
\theta$ is obtained from $N(E)=N_0 \mathrm{Re}(\cos \theta)$, where $N_0$ is
the Fermi level DOS in the normal state.

In order to calculate the current across a junction, the solution of the
Eqs. (\ref{eq-1}, \ref{eq-2}) for the proximity effect in the dirty SN
sandwich, including the influence of a clean 2DEG, has to be found. Eqs. (%
\ref{eq-1}, \ref{eq-2}) must be supplemented with the two sets of boundary
conditions: at the SN interface and at the 2DEG-N interface. For the SN
interface the boundary conditions are given by
\begin{eqnarray}
\gamma_{B_1} \xi_N \theta_N^\prime&=&\sin(\theta_S-\theta_N) , \\
\gamma_1 \xi_N \theta_N^\prime&=& \xi_S \theta_S^\prime ,  \label{eq-3}
\end{eqnarray}
where $\gamma_1 =\rho_S \xi_S/\rho_N \xi_N$ is a parameter characterizing
the strength of the proximity effect between S and N layers and $%
\gamma_{B_1} = R_B/\rho_N \xi_N$ describes the effect of the SN interface
transparency, $R_B$ is the SN interface resistance, $\rho_S$ and $\rho_N$
are the normal state resistivities. In the bulk of the superconductor, $%
\theta_S$ is given by $\theta_S =\arctan (i \Delta_0/E)$ where $\Delta_0$ is
the bulk value of the pair potential. For the N-2DEG interface, which
separates the disordered N layer (of thickness $d_N$) and the clean 2DEG,
the material-dependent parameters for the boundary conditions, similar to
Eqs. (3, 4), are found from the estimates \cite{Neurohr96}: $\gamma_2 \simeq
(3 k_{2DEG}/k_N^2l_N)\xi_N/\xi_{2DEG}$ and $\gamma_{B_2}=Z^2$, where $%
k_{2DEG}$ and $k_N$ are the Fermi wave-vectors in the 2DEG and in the N
layer, $l_N$ is the mean free path in the N layer, and $Z$ is the barrier
strength in the BTK model. The estimations show that for our NbN/Au/2DEG
junctions one may set $\gamma_2=0$.

The complete self-consistent problem requires numerical simulations. The
self-consistent solution is then used to determine the Andreev and normal
reflection coefficients at the 2DEG-N interface and the current across a
junction \cite{Neurohr96,Golubov96}
\begin{eqnarray}
A&=& \frac{\left|\sin \theta_N(\epsilon,0)\right|^2}{\left| 1+2Z^2+\cos
\theta_N (\epsilon,0) \right|^2} \\
B&=& \frac{4Z^2(1+Z^2)}{\left|1+2Z^2+ \cos \theta_N(\epsilon,0) \right|^2} \\
I&=&\frac{e ^2k_N W}{\pi^2 \hbar} \int^{+\infty}_{-\infty}\left[
f_0(\epsilon+eV) - f_0(\epsilon)\right]\left(1+A-B \right) d \epsilon
\end{eqnarray}
where $f_0$ is the Fermi distribution function, $V$ is the voltage drop in
the junction and $W$ is the contact width. In Eqs. (5, 6), the function $%
\theta_N(\epsilon,0)$ is taken in the N region near the N-2DEG interface.

An essential feature of the model \cite{Neurohr96,Golubov96} is
the existence of a gap in the density of states $\mathrm{Re}(\cos
\theta)$ of the N layer which is in proximity to a superconductor.
This leads to the characteristic two-gap structures in the energy
dependence of Andreev and normal reflection coefficients and thus
in the voltage dependence of the resistance \cite{Merkt97}. Fig.~5
shows the corresponding calculated $dV/dI-V_{dc}$ curves for
NbN/Au/2DEG junctions. The parameters for the calculated curves
were chosen within a realistic range given by the characteristic
material constants
\cite{Plathner96,Gousev94,Pambianchi94,Nunoz00,Dikin02,
Schaepers03}. As can be seen from Fig.~5 and Fig.~3, the
calculated and measured curves show a rather good qualitative
agreement. However, we could not find a good quantitative fit to
the experimental curves. The main discrepancies are in the
detailed shape and in the amplitude of variations of the
differential resistance. A weaker dip in the experimental curve
than calculated within the model \cite{Neurohr96} can possibly be
explained by smearing due to the inhomogeneity of the junction
(i.e. variation of the interface quality over the width of the
junction).

We have also considered the model of diffusive contact when a
disordered 2DEG channel exists between the clean 2DEG and SN
electrode and a potential drop is distributed between the SN-2DEG
interface and the disordered region in the 2DEG. Properties of
disordered contacts have been first studied theoretically by
Artemenko, Volkov, and Zaitsev \cite{Artemenko79} and later in
Refs.
\cite{VolkovAllsopp96,GolubovWilhelm97,Nazarov96,Belzig99,Zaitsev90,Lambert91,Wees92,Takane92,
Beenakker92,VolkovKlapwijk92,VolkovZaitsev93,Volkov93,Hekking93,Beenakker94,MelsenBeenakker96,
Lesovik97,VolkovTakayanagi97,Bezuglyi00}. The transport
measurements on the S/N structures in the diffusive limit have
been carried out in \cite{Kastalsky91,Magnee94,Quirion02} and have
revealed a conductance peak around zero bias voltage which arises
due to an interplay between Andreev scattering and
disorder-induced scattering in the normal electrode. The fitting
parameters we obtained within the model of diffusive contact (in
particular, large thickness of the disordered 2DEG channel $%
d_{N^{\prime}}\simeq~ 28\xi_{N^{\prime}}$) appeared to be unrealistic.
Therefore, it is less likely that the diffusive model can be applied in our
case when the 2DEG is in the extreme clean limit. We believe that the
observation of the two-gap like structure in the differential resistance is
related to the transport between the NS bilayer and clean 2DEG through the
clean constriction (the situation which was considered in \cite%
{Neurohr96,Golubov96}).

In conclusion, we fabricated highly transparent
superconductor/normal metal/two-dimensional electron gas junctions
formed by a superconducting NbN electrode, a thin (10nm) Au
interlayer, and a 2DEG in a InGaAs/InP heterostructure. High
junction transparency has been achieved by exploiting of a newly
developed process of Au/NbN evaporation and rapid annealing at
400$^{\circ }$C. A decrease in the differential resistance with
pronounced double dip structure was observed within the
superconducting energy gap in junctions investigated and its
magnetic field dependence was measured. It has been found that the
reduced subgap resistance remains in high magnetic fields.
Experimental data in zero field are analyzed within a model based
on the quasiclassical Green-function approach. The present results
suggest that our novel preparation method is not only advantageous
for the fabrication of highly transparent S/2DEG interfaces but it
might also have important implications for S/2DEG/S Josephson
junction based devices.

 The authors thank U. Z\"ulicke for fruitful
discussions, A.~van der Hart for performing electron beam
lithography on our samples and H. Kertz for his excellent
assistance during the measurements. This work was supported by the
Deutsche Forschungsgemeinschaft (DFG).

\newpage

\begin{figure}[t]
\vspace{1.5cm}
\includegraphics[width=9cm,angle=0]{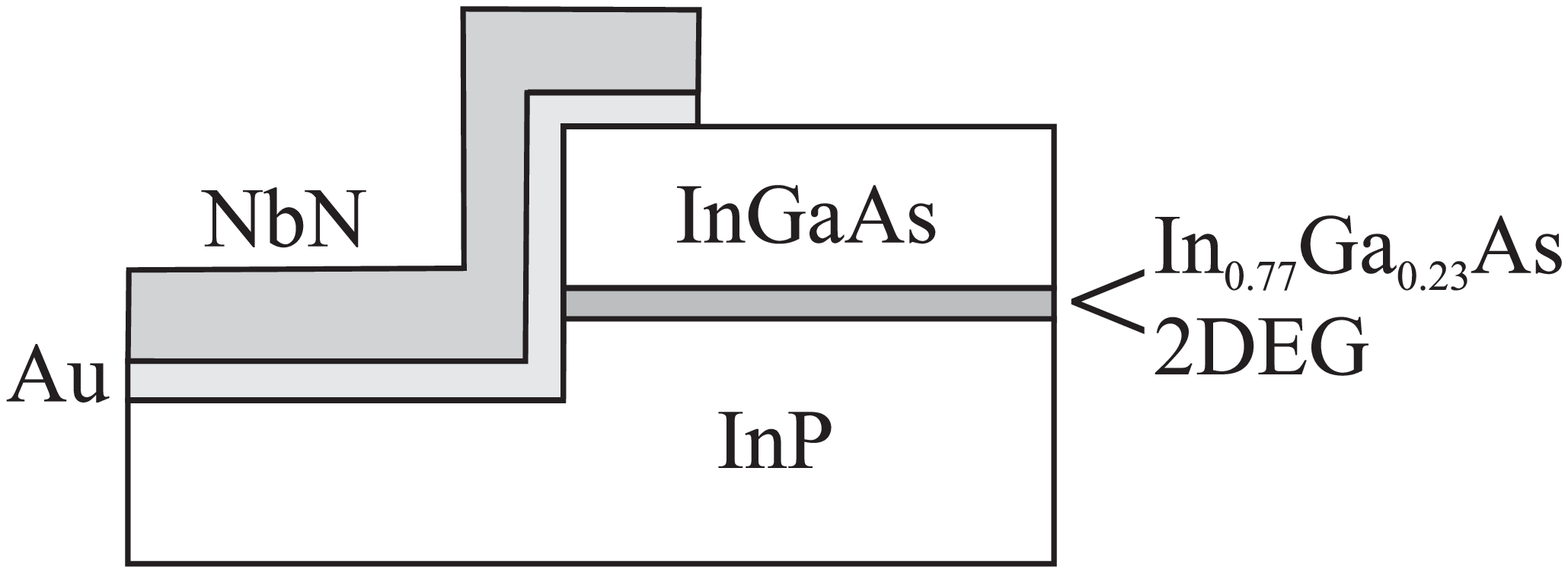}
\caption{Schematic of the sample layout. The two-dimensional
electron gas is located in the strained In$_{0.77}$Ga$_{0.23}$As
layer. The semiconductor mesa is defined by reactive ion etching.
The Au/NbN electrodes make contact at the sidewalls of the mesa. }
\label{Fig1}
\end{figure}

\begin{figure}[t]
\vspace{2cm}
\includegraphics[width=12cm,angle=0]{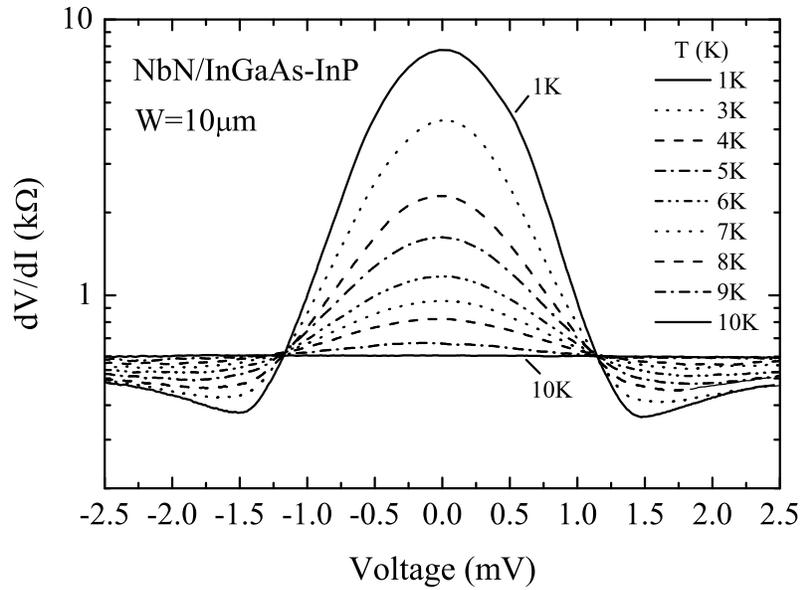}
\caption{Differential resistance vs bias voltage for a NbN/2DEG
interface at different temperatures.} \label{Fig2}
\end{figure}

\begin{figure}[t]
\includegraphics[width=12cm,angle=0]{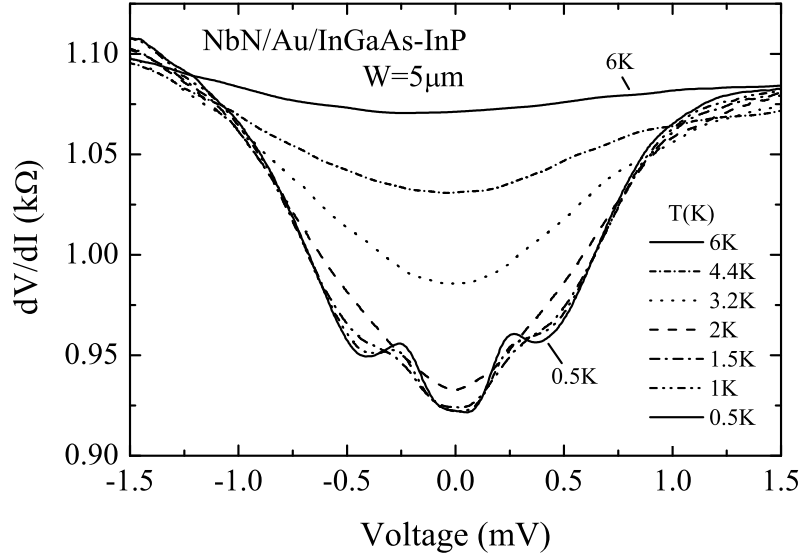}
\caption{Differential resistance as a function of the bias voltage for a
NbN/Au/2DEG interface. The measurements are performed at several
temperatures in the 0.5-6~K range.}
\label{Fig3}
\end{figure}

\begin{figure}[t]
\includegraphics[width=12cm,angle=0]{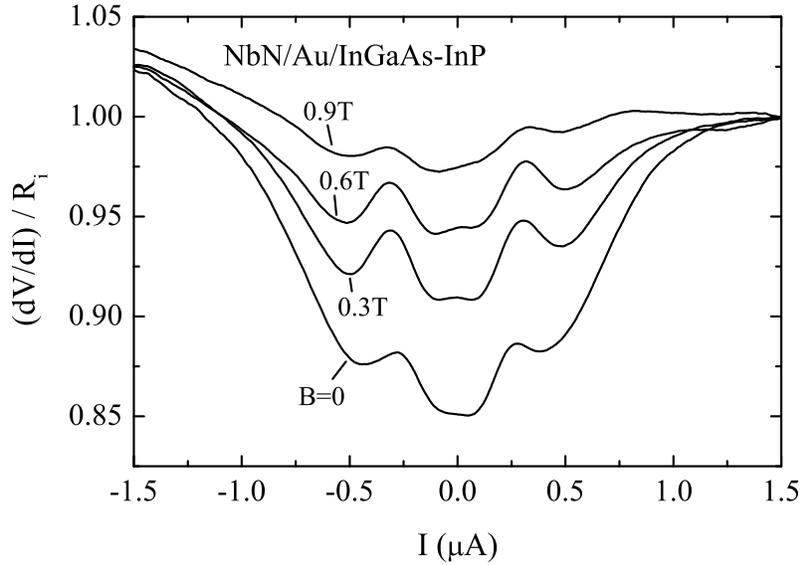}
\caption{Normalized differential resistance vs bias current of
NbN/Au/2DEG interface for several magnetic fields. $T=0.5$~K.}
\label{Fig4}
\end{figure}

\begin{figure}[t]
\includegraphics[width=12cm,angle=0]{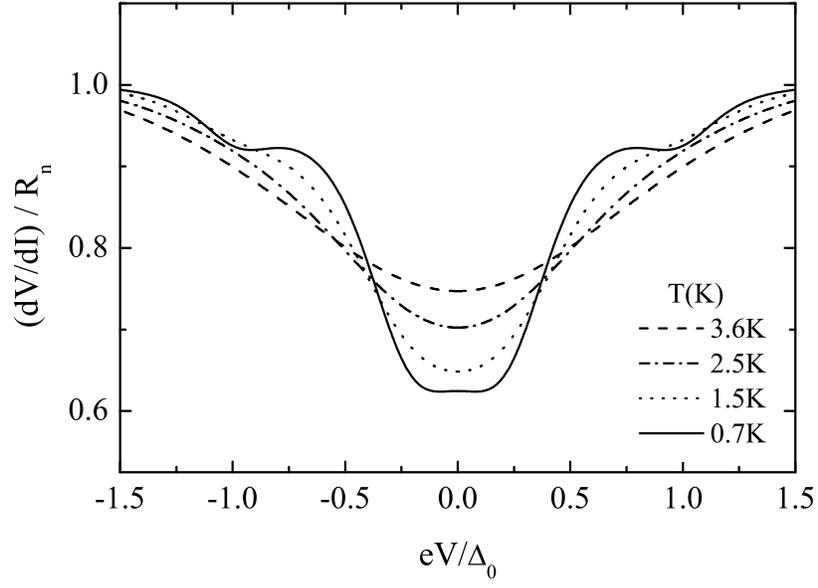}
\caption{ Calculated normalized differential resistance as a
function of the bias voltage according to the model
\cite{Neurohr96}. Parameters of the
model \cite{Neurohr96} are: $\gamma_1d_N/\xi_N=0.25$; $Z=0.35$; $%
\gamma_{B_1}d_N/\xi_N=5$} \label{Fig5}
\end{figure}

\end{document}